\documentclass[aps,prc,groupedaddress,showpacs,showkeys]{revtex4}

\usepackage{bm}
\usepackage{longtable}
\usepackage[dvips]{graphics}
\newcommand{\gs}{\raisebox{-.5ex}{ $\stackrel{>}{\scriptstyle \sim}$ }}
\newcommand{\ket}[1]{| #1 \rangle}
\newcommand{\bra}[1]{\langle #1 |}
\newcommand{\Ket}[1]{|| #1 \rangle}
\newcommand{\Bra}[1]{\langle #1 ||}
\newcommand{\Mass}{\mathrm{M}}
\newcommand{\mass}{\mathrm{m}}

\newcommand{\ie}{\emph{i.e.}}
\newcommand{\phf}{\mathrm{p.h.f.}}
\newcommand{\roundbra}[1]{\left( #1 \right|}
\newcommand{\roundket}[1]{\left| #1 \right)}
\newcommand{\roundbraket}[2]{\left(\,#1\,|\,#2\,\right)}

\newcommand{\Ref}[1]{(\ref{#1})}
\newcommand{\row}[4]{#1&&#2&&#3&&#4}
\newcommand{\twlvj }[3]{\left\{\!\begin{array}{cccccccc}
#1&\\&#2\\#3&\end{array}\!\right\}}
\newcommand{\jp}{\mathsf{j}_{p}}
\newcommand{\jL}{\mathsf{j}_{\Lambda}}

\newcommand{\nn}{{\nonumber}}

\newcommand{\br}{\begin{eqnarray}}
\newcommand{\er}{\end{eqnarray}}

\newcommand{\x}{\times }
\newcommand{\ninj }[9]{\left\{\negthinspace\begin{array}{ccc}
#1&#2&#3\\#4&#5&#6\\#7&#8&#9\end{array}\right\}}
\newcommand{\sixj }[6]{\left\{\negthinspace\begin{array}{ccc}
#1&#2&#3\\#4&#5&#6\end{array}\right\}}

\newcommand{\be}{\begin{equation}}
\newcommand{\ee}{\end{equation}}
\newcommand{\M }{{{\cal M}}}
\newcommand{\sqi}{\frac{1}{\sqrt{2}}}

\begin{document}
\title{Asymmetry parameter for nonmesonic hypernuclear decay}
\author{Cesar Barbero}
\affiliation{
Departamento de F\'{\i}sica,
Facultad de Ciencias Exactas, \\
Universidad Nacional de La Plata,
C.C. 67, 1900 La Plata, Argentina.
}
\author{Alfredo P. Gale\~ao}
\affiliation{
Instituto de F\'{\i}sica Te\'orica,
Universidade Estadual Paulista, \\
Rua Pamplona 145,
01405-900 S\~ao Paulo, SP, Brazil.
}
\author{Francisco Krmpoti\'c}
\affiliation{
Instituto de F\'{\i}sica,
Universidade de S\~ao Paulo, \\
C. P. 66318,
05315-970 S\~ao Paulo, SP, Brazil, \\
Instituto de F\'{\i}sica La Plata, CONICET,
1900 La Plata, Argentina, and\\
Facultad de Ciencias Astron\'omicas y Geof\'{\i}sicas, 
Universidad Nacional de La Plata, 1900 La Plata, Argentina.
}
\date{\today}

\begin{abstract}
We give general expressions for the vector asymmetry in the angular distribution of protons in the nonmesonic weak decay of polarized hypernuclei. From these we derive an explicit expression for the calculation of the asymmetry parameter, $a_\Lambda$, which is applicable to the specific cases of $^5_\Lambda$He and $^{12}_{\phantom{1}\Lambda}$C  described within the extreme shell model.
In contrast to the approximate formula widely used in the literature, it includes the effects of three-body kinematics in the final states of the decay and correctly treats the contribution of transitions originating from single-proton states beyond the $s$-shell.
This expression is then used for the corresponding numerical computation 
of $a_\Lambda$ within several one-meson-exchange models.
Besides the strictly local approximation usually adopted for the transition potential, we also consider the addition of the first-order nonlocality terms.
We find values for $a_\Lambda$ ranging from $-0.62$ to $-0.24$, in qualitative agreement with other theoretical estimates but in contradiction with some recent experimental determinations.
\end{abstract}


\maketitle

\section{Introduction \label{int}}

While the $\Lambda$ hyperon, in free space, decays 99.7\% of the time through the mesonic mode, $\Lambda \rightarrow \pi N$, inside nuclei this is Pauli-blocked, and,
already for $A \gs 5$, the weak decay is gradually dominated by the nonmesonic channel, $\Lambda N \rightarrow N N$, where the large momentum transfers involved
($\approx 400$ MeV/c) put the two emitted nucleons above the Fermi surface.
This decay mode is interesting since it offers a unique opportunity to probe the strangeness-changing weak interaction between hadrons. For a recent review of hypernuclear decay, see Ref.~\cite{Al02}.

For a long time, the experimental data for this process was restricted to the full nonmesonic decay rate, $\Gamma_{nm}$, and, in some cases, also the partial ones, $\Gamma_n=\Gamma(\Lambda n \to nn)$ and $\Gamma_p=\Gamma(\Lambda p \to np)$.
More recently, the first results, obtained at KEK \cite{Aj92,Aj00,Ma05}, for another important observable of nonmesonic decay, namely, the intrinsic asymmetry parameter, $a_\Lambda$, are becoming available.
This is  experimentally more demanding, as it requires measuring the asymmetry in the angular distribution of protons emitted in the decay of polarized hypernuclei. On the theoretical side, however, $a_\Lambda$ carries important new information, since it is determined by the interference terms between the parity-conserving (PC) and the parity-violating (PV) proton-induced transitions to final states with different isospins. In opposition to that, the decay rates depend only on the square moduli of the separate  components of the transition potential, being dominated by the PC ones. One expects, therefore, that the asymmetry parameter, besides being more sensitive to the PV amplitudes, will  have more discriminating power to constrain the proposed
mechanisms for nonmesonic hypernuclear decay.

Most of the theoretical work on this decay mode constructs the transition potential by means of one-meson-exchange (OME) models, the most complete ones including up to the whole ground pseudoscalar and vector meson octets
($\pi$, $\eta$, $K$, $\rho$, $\omega$, $K^*$) \cite{Du96,Pa97,Ba02}.
Recently we have extended such models to take into account the kinematical corrections due to the difference between the lambda and nucleon masses and the first-order nonlocality terms \cite{Ba03}.
There are also OME models that consider additional effects, such as correlated-two-pion exchange \cite{It02} and direct-quark interaction \cite{Sa02}.
In all these cases, to which we will refer below as strict OME models, the weak coupling constants for the pion are empirically determined from the free mesonic decay, and those of the remaining mesons by means of unitary-symmetry arguments \cite{Du96,Pa97}.

All such models reproduce quite easily the total nonmesonic decay rate, $\Gamma_{nm} = \Gamma_n + \Gamma_p$, but seem to strongly underestimate the experimental values for the $n/p$ branching ratio, $\Gamma_n/\Gamma_p$ \cite{Al02}. However, there are recent  indications, based on the intranuclear cascade model, that this might be due to contamination of the data by secondary nucleons unleashed by final state interactions (FSI) while the primary ones are traversing the residual nucleus \cite{Ga04}. Another serious discrepancy between theory and experiment in nonmesonic decay concerns the asymmetry parameter. The measurements favor a negative value for $^{12}_{\phantom{1}\Lambda}\mathrm{C}$ and a positive value for $^5_\Lambda\mathrm{He}$.
However, all existing calculations based on strict OME models
\cite{Du96,Pa97,Ba03,It02,Sa02} find values for $a_\Lambda$ between $-0.73$ and $-0.19$ \cite{Al02,Al05}.
Also, when results for $^{12}_{\phantom{1}\Lambda}\mathrm{C}$ are available in the same model, they are very similar to those for $^5_\Lambda\mathrm{He}$, as
expected, since the intrinsic asymmetry parameter, $a_\Lambda$, has been defined \cite{Ra92} in such a way as to subdue its dependence on the particular hypernucleus considered.
A recent attempt \cite{Al05} to explain this discrepancy along similar lines to those used for the $n/p$ problem has failed. As might be expected, the FSI do have an effect
in attenuating the asymmetry, but show no tendency to reverse its sign.
The only theoretical calculations that attain some agreement with the experimental data for $a_\Lambda$ are a first application of effective field theory to nonmesonic decay \cite{Pa04} and a very recent extension of the direct-quark interaction model to include sigma-meson exchange \cite{Sa05}. However, in both cases, one or more coupling constants are specifically adjusted to reproduce the experimental value of $a_\Lambda$ for $^5_\Lambda\mathrm{He}$.

Most calculations of the asymmetry parameter make use of an approximate formula (Eq.~\Ref{asym}, below) which, however, is valid only for $s$-shell hypernuclei. Since an essential aspect in the asymmetry puzzle presented above concerns the comparison of its values for $^5_\Lambda\mathrm{He}$ and $^{12}_{\phantom{1}\Lambda}\mathrm{C}$, it would be of great interest to have a simple expression that is applicable to both cases. This is the main objective of the present paper, in which a general formalism for the asymmetry parameter in nonmesonic decay is derived and subsequently particularized to these two hypernuclei. We start by presenting, in Section~\ref{asy}, the main steps in the derivation of the general expression, Eq.~\Ref{asymmetry}, of the vector hypernuclear asymmetry in terms of decay strengths. This is equivalent to Eq.~(27) of Ref.~\cite{Ra92}. However, we deviate considerably from that reference
from this point onwards. The main difference is that we do not make use of spectroscopic factors, but rather rely on spectroscopic amplitudes, which can then be computed in the nuclear structure model of choice. This has, in our view, two great advantages. Firstly, the spectroscopic amplitudes can be determined without any ambiguity as to their phases. This is particularly important for the asymmetry parameter, where, differently from the case of the decay rates, one is dealing with an interference phenomenon. Secondly, since, due to the large value of the momentum transferred in the fundamental process, nonmesonic decay is not significantly affected by details of nuclear structure, one can choose to work in the extreme shell model. Doing this, much of the summation over the final states of the residual nucleus can be explicitly performed, leading to very simple  expressions for the asymmetry. (See, for instance, Eq.~\Ref{omegas1}.)
The scheme for computing the decay strengths by means of an integration over the available phase space is presented in Section~\ref{strengths}, and the summations needed for the asymmetry parameter are performed in Section~\ref{asympar}.
Finally, the numerical results obtained by applying this formalism to the calculation of $a_\Lambda$ for, both $^5_\Lambda\mathrm{He}$, and $^{12}_{\phantom{1}\Lambda}\mathrm{C}$, in several strict OME models, are presented and discussed in Section~\ref{numerical}, where we also summarize our main conclusions.
Details of the derivation of the final expression for $a_\Lambda$ are given in Appendices \ref{K2}--\ref{K11}, and some identities that have been used for this purpose are listed in Appendix~\ref{12j}.

\section{Vector hypernuclear asymmetry \label{asy}}

Single-$\Lambda$ hypernuclei produced in a $(\pi^+,K^+)$ reaction, under favorable kinematical conditions,  are known to end up with considerable vector polarization
along the direction normal to the reaction plane,
$\hat{\bm{n}} = (\bm{p}_{\pi^+} \times \bm{p}_{K^+})/
|\bm{p}_{\pi^+} \times \bm{p}_{K^+}|$,
of which they retain a significant amount, $P_V$, even after they have cascaded down to their ground states by electromagnetic and strong processes \cite{Ej87,Aj92,Aj00}. Therefore, the initial \emph{mixed} state from which the hypernucleus will decay weakly can be described by the density matrix \cite[Eq.(9.29)]{Au70}
\begin{equation} \label{mixture}
\rho(J_I) = \frac{1}{2J_I+1}\left[ 1 + \frac{3}{J_I+1} P_V \bm{J}_I
\cdot \hat{\bm{n}} \right] \,,
\end{equation}
where $J_I$ is the hypernuclear spin.

The angular distribution of protons emitted in the proton-induced nonmesonic decay, $\Lambda\, p \to n\,p$, of the \emph{pure} initial hypernuclear state $\ket{J_IM_I}$ is given by Fermi's golden rule as
\begin{equation}\label{pure}
\frac{d\Gamma(J_IM_I \to \hat{\bm{p}}_2t_p)}{d\Omega_{p_2}} =
\int d\Omega_{p_1} \int dF\,
\sum_{s_1s_2M_F} \left|
\bra{\bm{p}_1 s_1 t_n\, \bm{p}_2 s_2 t_p\, \nu_FJ_FM_F} V \ket{J_IM_I}
\right|^2 \,.
\end{equation}
Here, $\bm{p}_1 s_1$ and $\bm{p}_2 s_2$ are the momenta and spin projections of the emitted neutron and proton, respectively, and we have introduced the compact notation ($\hbar = c =1$)%
\begin{equation} \label{compact}
\int dF \dots \;=\; 2\pi\, \sum_{\nu_FJ_F}
\int \frac{p_2^2\, dp_2}{(2\pi)^3} \int \frac{p_1^2\, dp_1}{(2\pi)^3}\;
\delta\left( \frac{p_1^2}{2\Mass} + \frac{p_2^2}{2\Mass}
+ \frac{|\bm{p}_1 + \bm{p}_2|^2}{2\Mass_F}
- \Delta_{\nu_FJ_F} \right) \dots \;,
\end{equation}
$\Mass$ being the nucleon mass;
$\Mass_F$, that of of the  residual nucleus, which is left in state
$\ket{\nu_FJ_FM_F}$ where $\nu_F$ specifies the remaining quantum numbers besides those related to the nuclear spin; and $\Delta_{\nu_FJ_F}$, the liberated energy.
(To avoid confusion, we will be using Roman font ($\Mass$,$\mass$) for masses
and italic font ($M$,$m$) for azimuthal quantum numbers.)
Also indicated in Eq.~\Ref{pure} are the isospin projections $t_n\equiv -1/2$ and 
$t_p\equiv +1/2$ of the neutron and proton, respectively.
The transition amplitude includes both the direct and the exchange contributions, \ie,
\begin{eqnarray}
\lefteqn{
\bra{\bm{p}_1 s_1 t_n\, \bm{p}_2 s_2 t_p\, \nu_FJ_FM_F} V \ket{J_IM_I}
}
\nonumber \\ &=&
\roundbra{\bm{p}_1 s_1 t_n\, \bm{p}_2 s_2 t_p\, \nu_FJ_FM_F} V \ket{J_IM_I}
-
\roundbra{\bm{p}_2 s_2 t_p\, \bm{p}_1 s_1 t_n\, \nu_FJ_FM_F} V \ket{J_IM_I}
\,,
\label{npamplitude}
\end{eqnarray}
where the round bras stand for simple (nonantisymmetrized) product states for the emitted nucleons and the transition potential, $V$, is extracted from the Feynman amplitude for the direct process \cite{Ba03}.

\begin{figure}
\scalebox{.75}{
\includegraphics*[125pt,210pt][485pt,675pt]{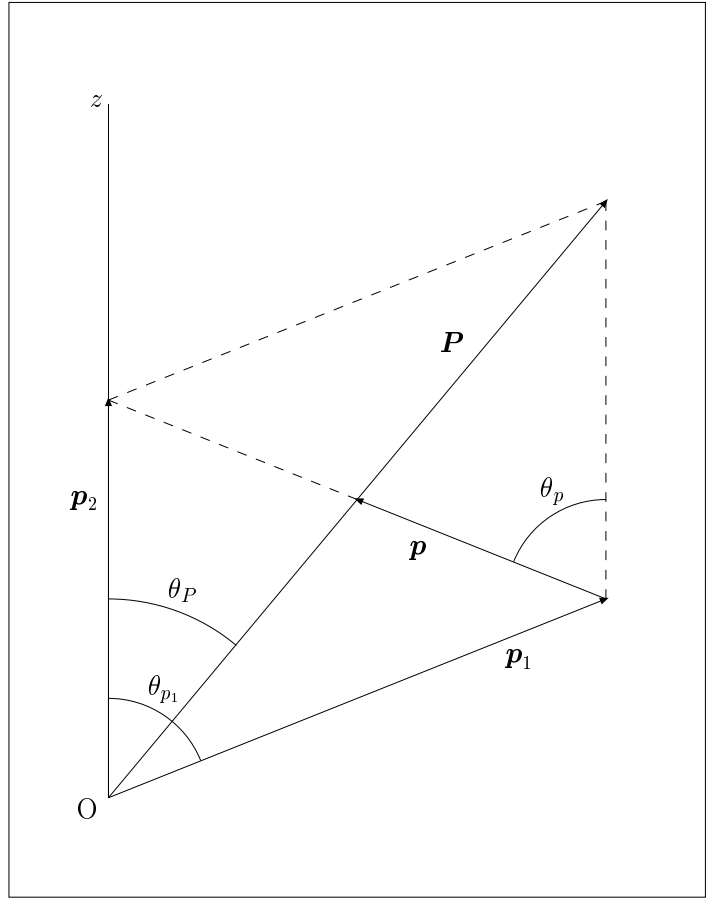}
}
\caption{Coordinate system for the calculation of decay strengths. \label{helicity}}
\end{figure}

It is then possible to show \cite{Ra92,Ga04a}, by taking the appropriate average of Eq.~\Ref{pure}, that the angular distribution of
protons from the decay of the polarized mixed state described by
Eq.~\Ref{mixture} has the form
\begin{equation}\label{angular}
\frac{d\Gamma[ \rho(J_I) \to \hat{\bm{p}}_2 t_p ]}{d\Omega_{p_2}} =
\frac{\Gamma_p}{4\pi}
\, \left( 1 + P_V A_V\, \hat{\bm{p}}_2 \cdot \hat{\bm{n}} \right) \,,
\end{equation}
where $\Gamma_p$ is the full proton-induced decay rate,
and the \emph{vector hypernuclear asymmetry}, $A_V$, is given by
\begin{equation}\label{asymmetry}
A_V = \frac{3}{J_I+1}\; \frac{\sum_{M_I} M_I \sigma( J_I M_I)}
{\sum_{M_I}  \sigma( J_I M_I)} \,.
\end{equation}
The new quantities introduced above are the \emph{decay strengths},
\begin{equation}
\sigma( J_I M_I) \;=\; \int d\Omega_{p_1} \int dF
\sum_{s_1 s_2 M_F} \left|
\bra{\bm{p}_1 s_1 t_n\, \bm{p}_2 s_2 t_p\, \nu_FJ_FM_F} V
\ket{J_IM_I}_\phf
\right|^2  \,,
\label{strength}
\end{equation}
where the subscript $\phf$ indicates that one is dealing here with the transition amplitude in the proton helicity frame, in which the direction for angular momentum quantization is that of the proton momentum.
Equivalently, one can choose, for the calculation of the decay strengths, a coordinate system having the $z$-axis pointing along the proton momentum, and proceed as usually.
This is depicted in Fig.~\ref{helicity}.

%

It is clear that, with the help of Eq.~\Ref{angular}, one can extract the value
of the product $P_VA_V$ from the counting rates parallel and opposite to the
polarization direction, by taking the ratio of their difference to their sum. Assuming that $P_V$ can also be independently measured, or calculated, this  experimentally determines the vector hypernuclear asymmetry, $A_V$.

\section{Decay strengths \label{strengths}}

To compute the decay strengths, it is convenient to rewrite the transition amplitudes in Eq.~\Ref{strength} in the total spin $(S,M_S)$ and isospin $(T,M_T)$ basis. 
We start from the relation  
\begin{eqnarray}
\lefteqn{
\roundket{\bm{p}_1 s_1 t_n\, \bm{p}_2 s_2 t_p} 
-
\roundket{\bm{p}_2 s_2 t_p \, \bm{p}_1 s_1 t_n}
}
\nonumber \\ &=&
\sum_{SM_STM_T} (1/2\, s_1\; 1/2\, s_2|SM_S) \, (1/2\, t_n \; 1/2\, t_p|TM_T) 
\nonumber \\ &\times&
\left[\,
\roundket{\bm{p}  \bm{P} SM_STM_T} 
- (-)^{S+T} 
\roundket{\bm{-p}  \bm{P} SM_STM_T}
\,\right] \,,
\label{transformation}
\end{eqnarray}
where we have also changed the representation to relative and total momenta,
given respectively by
\begin{eqnarray}
\bm{p} &=& \frac{1}{2} (\bm{p}_2 - \bm{p}_1) \,,
\nn\\
\bm{P} &=& \bm{p}_1 + \bm{p}_2 \,.
\label{momenta}
\end{eqnarray}
Since we are taking $t_n\equiv-1/2$ and $t_p\equiv+1/2$,  we can write
\begin{equation}
(1/2\, t_n \; 1/2\, t_p|TM_T) = 
\frac{1}{\sqrt{2}}\, \left( \delta_{T1} - \delta_{T0} \right) \, \delta_{M_T0} \,,
\end{equation}
and performing the transformation \Ref{transformation} in Eq.~\Ref{npamplitude}, we get   
\begin{eqnarray}
\lefteqn{
\bra{\bm{p}_1 s_1 t_n\, \bm{p}_2 s_2 t_p\, \nu_FJ_FM_F} V \ket{J_IM_I}
}
\nonumber \\ &=&
- \sum_{SM_S} (1/2\, s_1\; 1/2\, s_2|SM_S) \, 
\sum_{T} \, (-)^{T} \, \bra{\bm{p}  \bm{P} SM_ST\, \nu_F J_FM_F} V \ket{ J_IM_I} \,,
\label{transformation1}
\end{eqnarray}
where we have defined 
\begin{eqnarray}
\lefteqn{
\bra{\bm{p}  \bm{P} SM_ST\, \nu_F J_FM_F} V \ket{ J_IM_I}
}
\nonumber \\ &=&
\frac{1}{\sqrt{2}}
\left[
\roundbra{\bm{p}  \bm{P} SM_ST\, \nu_F J_FM_F} V \ket{ J_IM_I}
- (-)^{S+T}
\roundbra{\bm{-p}  \bm{P} SM_ST\, \nu_F J_FM_F} V \ket{ J_IM_I}
\right] \,,
\label{isoamplitude}
\end{eqnarray}
 dropping, for simplicity, the $M_T=0$ labels, as shall be done henceforth. 
Finally, introducing Eq.~\Ref{transformation1} into Eq.~\Ref{strength}, and making use of the orthogonality of the Clebsch-Gordan coefficients in spin space,  we are left with
\begin{equation}\label{strength1}
\sigma( J_I M_I) =
\int d\Omega_{p_1} \int dF\,
\sum_{SM_SM_F} \left|\sum_{T}(-)^T
\bra{\bm{p}  \bm{P} SM_ST\, \nu_F J_FM_F} V \ket{ J_IM_I}_\phf
\right|^2 \,.
\end{equation}

For the integration in Eq.~\Ref{strength1}, there are 6 momentum variables involved, namely, the components of $\bm{p}_1$ and $\bm{p}_2$. These, however, are not all independent.
The choice of $z$-axis in Fig.~\ref{helicity} eliminates two angular variables. Also, the energy conservation condition in Eq.~\Ref{compact} gives one relation to be satisfied.
This leaves $6-3=3$ independent variables. A  convenient choice is
\begin{equation} \label{independent}
\mbox{independent variables:} \quad p_2\,, \theta_{p_1} \,, \phi_{p_1} \,.
\end{equation}
Simple trigonometry, applied in Fig.~\ref{helicity}, leads to the relations
\begin{eqnarray}
4p^2 &=& p_1^2 + p_2^2 - 2 p_1 p_2 \cos \theta_{p_1} \,,
\nonumber \\
P^2 &=& p_1^2 + p_2^2 + 2 p_1 p_2 \cos \theta_{p_1} \,,
\nonumber \\
\cos \theta_p &=& \frac{p_2 - p_1\cos\theta_{p_1}}{2p} \,,
\nonumber \\
\cos \theta_P &=&  \frac{p_2 + p_1\cos\theta_{p_1}}{P} \,,
\label{variables}
\end{eqnarray}
which, together with the energy conservation condition, determine all momentum variables in terms of the set in Eq.~\Ref{independent}.
Notice that the azimuthal angles of the several momenta are
related as follows
\begin{equation}\label{azimuthal}
\phi_p = \phi_{p_1} + \pi\,,  \quad   \phi_P = \phi_{p_1} \,.
\end{equation}

For the transition amplitude, we expand the final state in terms of the relative and center-of-mass partial waves of the emitted nucleons  \cite[(2.5)]{Ba02}, getting,
\begin{eqnarray}
\lefteqn{
\bra{\bm{p}  \bm{P} SM_ST\, \nu_F J_FM_F} V \ket{ J_IM_I}_\phf
}
\nonumber \\ &=&
(4\pi)^2 \sum_{lL\lambda J} i^{-l-L}\,
[Y_{l}(\theta_p,\phi_{p_1}+\pi) \otimes Y_{L}(\theta_P,\phi_{p_1} )]_{\lambda\mu}
\nonumber \\
&\times& (\lambda \mu SM_S|JM_J)(JM_JJ_FM_F|J_IM_I)
\bra{plPL\lambda SJT\nu_FJ_F;J_I} V \ket{ J_I},
\label{multipole}
\end{eqnarray}
where the values of $\mu$ and $M_J$ are fixed by the relations  $M_I=M_J+M_F=\mu+M_S+M_F$.
Due to the rotational invariance of $V$, the last matrix element in Eq.~\Ref{multipole} is independent of $M_I$, and this label has, therefore, been omitted. For the same reason, the subscript $\phf$ has also been dropped.
Notice that, from Eq.~\Ref{isoamplitude} and the well known behavior of the spherical harmonics under parity, one has
\begin{equation}
\bra{plPL\lambda SJT\nu_FJ_F;J_I} V \ket{ J_I}
\;=\;
\frac{1}{\sqrt{2}} \left[1 - (-)^{l+S+T}\right]
\roundbra{plPL\lambda SJT\nu_FJ_F;J_I} V \ket{ J_I} \,.
\end{equation}

Upon integration on the angle $\phi_{p_1}$, Eq.~\Ref{strength1} gives, then,
\begin{eqnarray}
\sigma(J_I M_I) &=& \frac{1}{2} (4\pi)^5
\int d\cos\theta_{p_1} \int dF\,
\sum_{SM_SM_F} \,\biggl|
\sum_{lL\lambda JT} (-)^T\, i^{-l-L}\,
[Y_{l}(\theta_p,\pi) \otimes Y_{L}(\theta_P,0)]_{\lambda\mu}
\nonumber \\ &\times&
(\lambda \mu SM_S|JM_J) (JM_JJ_FM_F|J_IM_I)
\bra{plPL\lambda SJT\nu_FJ_F;J_I}V\ket{J_I} \biggr|^2 \,.
\label{strength2}
\end{eqnarray}
It can be shown quite generally that
\begin{eqnarray}
\lefteqn{
[Y_{l}(\theta_p,\phi_p) \otimes Y_{L}(\theta_P,\phi_P)]_{\lambda\mu}\,
[Y_{l'}(\theta_p,\phi_p) \otimes Y_{L'}(\theta_P,\phi_P)]_{\lambda'\mu}^\ast
}
\nonumber \\
&=&
(4\pi)^{-1}\, (-)^{l'+L'}\,
\hat{l}\hat{l'}\hat{L}\hat{L'}\hat{\lambda'}
\nonumber \\ &\times& \sum_{kK\kappa}\hat{\kappa}
(l0l'0|k0)(L0L'0|K0)(\lambda'\mu \kappa 0|\lambda\mu)
\nn\\&\x&\ninj{l}{l'}{k}{L}{L'}{K}{\lambda}{\lambda'}{\kappa}
[Y_{k}(\theta_p,\phi_p)\otimes Y_{K}(\theta_P,\phi_P)] _{\kappa 0} \,,
\end{eqnarray}
where $\hat{l}=\sqrt{2l+1}$ and similarly for other angular momentum labels.
Therefore, upon opening the square and performing the summations on $M_S$ and $M_F$, Eq.~\Ref{strength2} becomes
\begin{eqnarray}
\sigma(J_I M_I)&=&\frac{1}{2} (4\pi)^{4} \int d\cos\theta_{p_1} \int dF\,
\sum_{STT'}(-)^{T+T'}
\nonumber \\ &\times&
\sum_{lL\lambda J} \; \sum_{l'L'\lambda' J'}
i^{-l'-L'-l-L}\;(-)^{\lambda+S+J+J'+J_I+J_F}\;
\hat{l}\hat{l}'\hat{L}\hat{L'}\hat{\lambda}\hat{\lambda'}\hat{J}\hat{J'}
\nonumber \\ &\times&
\sum_{kK\kappa} \;
\hat{\kappa}\hat{J}_{I} \,
(J_I M_I \kappa 0|J_IM_I)(l0l'0|k0)(L0L'0|K0) \,
[Y_{k}(\theta_p,\pi)\otimes Y_{K}(\theta_P,0)] _{\kappa 0}
\nonumber \\ &\times&
\sixj{J_I}{\kappa}{J_I}{J}{J_F}{J'}
\sixj{\kappa}{J'}{J}{S}{\lambda}{\lambda'}
\ninj{l}{l'}{k}{L}{L'}{K}{\lambda}{\lambda'}{\kappa}
\nonumber \\ &\times&
\bra{plPL\lambda SJT\nu_FJ_F;J_I}V\ket{J_I}
\bra{pl'PL'\lambda' SJ'T'\nu_FJ_F;J_I}V\ket{J_I}^\ast \,.
\label{strength3}
\end{eqnarray}

\section{Asymmetry parameter \label{asympar}}

In order to carry out the summations on $M_I$ needed in Eq.~\Ref{asymmetry}, we first rewrite it in the form
\begin{equation}\label{asymmetry1}
A_V = 3\; \sqrt{\frac{J_I}{J_I+1}}\; \frac{\sigma_1(J_I)}{\sigma_0(J_I)} \,,
\end{equation}
where we have introduced  the \emph{decay moments}
\begin{eqnarray}
\sigma_0(J_I) &=& \sum_{M_I}  \sigma( J_I M_I) \,,
\label{moment0}
\\
\sigma_1(J_I) &=& \frac{1}{\sqrt{J_I(J_I+1)}} \sum_{M_I} M_I \sigma( J_I M_I) \,.
\label{moment1}
\end{eqnarray}
Then we take advantage of the particular values
\begin{eqnarray}
(J_IM_I00|J_IM_I) &=& 1,
\nonumber \\
(J_IM_I10|J_IM_I) &=& M_I/\sqrt{J_I(J_I+1)},
\end{eqnarray}
and use the orthogonality relation
\begin{eqnarray}
\sum_{M_I}(J_IM_I\kappa 0|J_IM_I)(J_IM_I\kappa' 0|J_IM_I) = \delta_{\kappa\kappa'} \hat{J}_I ^{2} \hat{\kappa}^{-2}
\end{eqnarray}
to get, for $\kappa=0 \mbox{ and }1$,
\begin{eqnarray}
\sigma_\kappa(J_I) &=&\frac{1}{2} (4\pi)^{4}\, \hat{J}_I^3\, \hat{\kappa}^{-1}
\int d\cos\theta_{p_1} \int dF\,
\sum_{STT'}(-)^{T+T'}
\nonumber \\ &\times&
\sum_{lL\lambda J} \; \sum_{l'L'\lambda' J'}
i^{-l'-L'-l-L}\;(-)^{\lambda+S+J+J'+J_I+J_F}\;
\hat{l}\hat{l}'\hat{L}\hat{L'}\hat{\lambda}\hat{\lambda'}\hat{J}\hat{J'}
\nonumber \\ &\times&
\sum_{kK} \; (l0l'0|k0)(L0L'0|K0) \,
[Y_{k}(\theta_p,\pi)\otimes Y_{K}(\theta_P,0)] _{\kappa 0}
\nonumber \\ &\times&
\sixj{J_I}{\kappa}{J_I}{J}{J_F}{J'}
\sixj{\kappa}{J'}{J}{S}{\lambda}{\lambda'}
\ninj{l}{l'}{k}{L}{L'}{K}{\lambda}{\lambda'}{\kappa}
\nonumber \\ &\times&
\bra{plPL\lambda SJT\nu_FJ_F;J_I}V\ket{J_I}
\bra{pl'PL'\lambda' SJ'T'\nu_FJ_F;J_I}V\ket{J_I}^\ast \,.
\label{moments}
\end{eqnarray}

From  \cite[(2.13)]{Ba02} (see also \cite{Ga04b}):
\begin{eqnarray}
\lefteqn{
\bra{plPL\lambda SJT\nu_FJ_F;J_I}V\ket{J_I}
}
\nonumber \\ &=&
(-)^{J_F+J-J_I}\hat{J}_I^{-1} \sum_{\jp}
\Bra{J_I}\left(a_{\jL }^\dag a_{\jp}^\dag \right)_{J}\Ket{\nu_FJ_F}\,
\M(plPL\lambda SJT;{\jL \jp}) \,,
\label{decomposition}
\end{eqnarray}
where $\jL\equiv n_\Lambda\,l_\Lambda\,j_\Lambda$ and $\jp\equiv n_p\,l_p\,j_p$  are the single-particle states for the lambda and proton, respectively,  and
\begin{eqnarray}
\M(plPL\lambda SJT;\jL \jp)
&=&\sqi\left[1-(-)^{l+S+T}\right]
({plPL\lambda SJT}|V|{\jL \jp J}).
\label{twobody}
\end{eqnarray}
We are working in the weak-coupling model (WCM), where the hyperon is assumed to stay in the $\jL=1s_{1/2}$ single-particle state, and the initial hypernuclear state $\ket{J_I}$ is built by the simple coupling  of  this orbital to the core, taken as the $^{A-1}Z$ ground state $\ket{J_C}$, \ie,
$\ket{J_I}\equiv \ket{(\jL J_C)J_I}$.
Under these circumstances, the two-particle spectroscopic amplitudes in Eq.~\Ref{decomposition} are cast as \cite{Ba02,Kr03}
\begin{eqnarray}
\Bra{J_I}\left(a_{\jL }^\dag a_{\jp}^\dag \right)_{J}\Ket{\nu_FJ_F}
&=&(-)^{J+J_I+J_F}\hat{J}\hat{J}_I
\sixj{J_C}{J_I}{j_\Lambda}{J}{j_p}{J_F}
\Bra{J_C}a_{\jp}^\dag\Ket{\nu_FJ_F}.
\label{spectroscopic}
\end{eqnarray}

To continue, we will adopt the extreme shell model (ESM)
and restrict our attention to cases where the single-proton states are completely filled in $\ket{J_C}$. This is so, within the ESM, for the cores of, both $^5_\Lambda$He  $(J_I=1/2,J_C=0)$, and
$^{12}_{\phantom{1}\Lambda}$C $(J_I=1,J_C=3/2)$. In such cases, the final nuclear states take the form
$\ket{\nu_FJ_F} \equiv \ket{(\jp^{-1}J_C)J_F}$, and we can associate the extra label $\nu_F$ with the occupied single-proton states, $\jp$.
Consequently, on one hand, only one term contributes to the sum in Eq.~\Ref{decomposition}, and, on the other, the corresponding single-proton spectroscopic amplitude in Eq.~\Ref{spectroscopic} is  given by
 \begin{eqnarray}
\Bra{J_C}a_{\jp}^\dag\Ket{\nu_FJ_F}=(-)^{J_F+J_C+j_p}\hat{J}_F \,.
\label{spectroscopic1}
\end{eqnarray}
Notice also that, within this description, the liberated energies are independent of  $J_F$, \ie, $\Delta_{\nu_FJ_F} \to \Delta_{\jp}$. This suggests rewriting Eq.~\Ref{compact} as
\begin{equation} \label{compact1}
\int dF \dots \;=\; \frac{1}{(2\pi)^5} \sum_{\jp} \int dF_{\jp} \sum_{J_F} \dots \;,
\end{equation}
where
\begin{equation} \label{compact2}
\int dF_{\jp} \dots \;=\;
\int p_2^2\, dp_2 \int p_1^2\, dp_1\;
\delta\left( \frac{p_1^2}{2\Mass} + \frac{p_2^2}{2\Mass}
+ \frac{|\bm{p}_1 + \bm{p}_2|^2}{2\Mass_F}
- \Delta_{\jp} \right) \dots \;.
\end{equation}
Putting all this together and performing the summation on $J_F$, we finally get
\begin{eqnarray}
\sigma_\kappa(J_I) &=& \frac{4}{\pi}\, \hat{J}_I^3\, \hat{\kappa}^{-1}
\sum_{\jp} \int d\cos\theta_{p_1} \int dF_{\jp}
\sum_{STT'}(-)^{T+T'}
\nonumber \\ &\times&
\sum_{lL\lambda J} \; \sum_{l'L'\lambda' J'}
i^{-l'-L'-l-L}\; (-)^{\lambda+S+J_I+J_C-j_p+\kappa}\;
\hat{l}\hat{l}'\hat{L}\hat{L'}\hat{\lambda}\hat{\lambda'}\hat{J}^2\hat{J'}^2
\nonumber \\ &\times&
\sum_{kK} \; (l0l'0|k0)(L0L'0|K0) \,
[Y_{k}(\theta_p,\pi)\otimes Y_{K}(\theta_P,0)] _{\kappa 0}
\nonumber \\ &\times&
\sixj{j_\Lambda}{J_I}{J_C}{J_I}{j_\Lambda}{\kappa}
\sixj{\kappa}{j_\Lambda}{j_\Lambda}{j_p}{J}{J'}
\sixj{\kappa}{J'}{J}{S}{\lambda}{\lambda'}
\ninj{l}{l'}{k}{L}{L'}{K}{\lambda}{\lambda'}{\kappa}
\nonumber \\ &\times &
\mathcal{M}(plPL\lambda SJT;{\jL \jp})
\mathcal{M}^*(pl'PL'\lambda' SJ'T';{\jL \jp}).
\label{moments1}
\end{eqnarray}

From Eqs. \Ref{asymmetry1} and \Ref{moments1}, it can be seen that
all the dependence on $J_I$ is contained in the factor
\begin{eqnarray}
-\, \sqrt{\frac{3J_I}{J_I+1}} \;
\frac{\sixj{1/2}{J_I}{J_C}{J_I}{1/2}{1}}
{\sixj{1/2}{J_I}{J_C}{J_I}{1/2}{0}}
&=& \left\{
\begin{array}{ccc}
1
&\hspace{2em}\mbox{for}\hspace{2em}&J_I=J_C+1/2 \,, \\
-\frac{J_I}{J_I+1}
&\hspace{2em}\mbox{for}\hspace{2em}&J_I=J_C-1/2 \,.
\end{array}\right. 
\label{factor}
\end{eqnarray}
Thus, within the framework of the WCM, one frequently introduces the
intrinsic $\Lambda$ asymmetry parameter, $a_\Lambda$, \cite {Ra92,Al02} defined
as
\begin{eqnarray}
a_\Lambda&=&
\left\{
\begin{array}{ccc}
A_V&
\hspace{2em}\mbox{for}\hspace{2em}&J_I=J_C+1/2 \,, \\
-\frac{J_I+1}{J_I}A_V&
\hspace{2em}\mbox{for}\hspace{2em}&J_I=J_C-1/2 \,,
\end{array}\right. 
\label{parameter}
\end{eqnarray}
which does not depend on the hypernuclear spin, as we have just shown within the ESM and for core states having no open proton subshells. For such cases, we get
\begin{equation}\label{parameter1}
a_\Lambda = \frac{\omega_1}{\omega_0} \,,
\end{equation}
with, for $\kappa=0 \mbox{ and } 1$,
\begin{eqnarray}
\omega_\kappa &=&
8\sqrt{2} \,
\sum_{\jp} \int d\cos\theta_{p_1} \int dF_{\jp}
\sum_{STT'}(-)^{T+T'}
\nonumber \\ &\times&
\sum_{lL\lambda J} \; \sum_{l'L'\lambda' J'}
i^{-l'-L'-l-L}\; (-)^{\lambda+S+j_p+\frac{1}{2}}\;
\hat{l}\hat{l}'\hat{L}\hat{L'}\hat{\lambda}\hat{\lambda'}\hat{J}^2\hat{J'}^2
\nonumber \\ &\times&
\sum_{kK} \; (l0l'0|k0)(L0L'0|K0) \,
[Y_{k}(\theta_p,\pi)\otimes Y_{K}(\theta_P,0)] _{\kappa 0}
\nonumber \\ &\times&
\sixj{\kappa}{1/2}{1/2}{j_p}{J}{J'}
\sixj{\kappa}{J'}{J}{S}{\lambda}{\lambda'}
\ninj{l}{l'}{k}{L}{L'}{K}{\lambda}{\lambda'}{\kappa}
\nonumber \\ &\times &
\mathcal{M}(plPL\lambda SJT;{\jL \jp})
\mathcal{M}^*(pl'PL'\lambda' SJ'T';{\jL \jp}).
\label{omegas}
\end{eqnarray}
It can be shown \cite{Ga04a} that $\omega_0=\Gamma_p$. Therefore, the new information carried by $a_\Lambda$ comes from the numerator in Eq.~\Ref{parameter1}, \ie, from $\omega_1$.

The orbital angular momenta in Eq.~\Ref{omegas} obey the restrictions: 
\begin{eqnarray}
(-)^{l+l'-k} &=& +1 \,,
\nonumber 
\\
(-)^{L+L'-K} &=& +1 \,,
\nonumber 
\\
(-)^{k+K-\kappa} &=& +1 \,.
\label{restrictions} 
\end{eqnarray}
The first two follow from well known properties of Clebsch-Gordan coefficients with vanishing azimuthal quantum numbers. The last one can be obtained by first deriving the general relation
\begin{equation}
[Y_{k}(\theta_p,\phi_p)\otimes Y_{K}(\theta_P,\phi_P)]^\ast_{\kappa \mu} =
(-)^{k+K-\kappa+\mu}\,
[Y_{k}(\theta_p,\phi_p)\otimes Y_{K}(\theta_P,\phi_P)]_{\kappa \, -\!\mu} \,.
\end{equation}
Then, recalling that any spherical harmonic with azimuthal angle equal to, either $0$,  or $\pi$, is real, one gets,
\begin{equation}
[Y_{k}(\theta_p,\pi)\otimes Y_{K}(\theta_P,0)]_{\kappa \, 0} =
(-)^{k+K-\kappa}\,
[Y_{k}(\theta_p,\pi)\otimes Y_{K}(\theta_P,0)]_{\kappa \, 0} \,,
\end{equation}
from which the third one of Eqs.~\Ref{restrictions} follows immediately.
The presence of the phase factor $i^{-l'-L'-l-L}$ in Eq.~\Ref{omegas} may seem disquieting at first sight. However, by taking the complex conjugate of that equation, interchanging the dummy variables $lL\lambda JT \leftrightarrow  l'L'\lambda' J'T'$, and making use of Eqs.~\Ref{restrictions} and of the symmetry properties of angular-momentum coupling and recoupling coefficients, one easily gets the relation $\omega_\kappa^\ast = \omega_\kappa$, showing that these quantities are real, as they should be by definition.

To compute the two-body matrix elements defined in Eq.~\Ref{twobody}, we resort to a Moshinsky transformation \cite{Mo59} of the initial state, and phenomenologically add initial and final short-range correlations.
(For more detail on this and related points, see Refs. \cite{Ba02} and \cite{Ba03}.)
For  $^{5}_\Lambda$He, the sole contribution to Eq.~\Ref{omegas} comes from the $1s_{1/2}$ proton state, and one can put $L=L'=K=0$.  
On the other hand, for $^{12}_{\phantom{1}\Lambda }$C, also the $1p_{3/2}$ state contributes, in which case $L$ and $L'$ can each take the values $0$ and $1$. Consequently one could, in principle, have $K=0,1$ and $2$ in Eq.~\Ref{omegas}. 
But we prove in Appendix~\ref{K2} that the contribution with $K=2$ vanishes identically, both for $\kappa=0$, and for $\kappa=1$. Similarly, we prove in Appendix~\ref{K10} that the contribution with $K=1$ vanishes for $\kappa=0$. We do not have an analytical proof that the contribution with $K=\kappa=1$ vanishes, but we show in Appendix~\ref{K11} that it is, in any case, negligibly small. 
Therefore, only the term with $K=0$ survives in Eq.~\Ref{omegas}, and it reduces to the following expression, that can be used for the two hypernuclei:
\begin{eqnarray}
\omega_\kappa &=&
(-)^\kappa \, \frac{8}{\sqrt{2\pi}} \, \hat{\kappa}^{-1} \,
\sum_{\jp} \int d\cos\theta_{p_1} \int dF_{\jp} \,
Y_{\kappa 0}(\theta_p,0)
\nonumber \\ &\times&
\sum_{TT'}(-)^{T+T'}
\sum_{LS} \; \sum_{l\lambda J} \; \sum_{l'\lambda' J'}
i^{l-l'}\; (-)^{\lambda+\lambda'+S+L+j_p+\frac{1}{2}}
\nonumber \\ &\times&
\hat{l}\hat{l}'\hat{\lambda}\hat{\lambda'}\hat{J}^2\hat{J'}^2 \,
(l0l'0|\kappa 0) \,
\nonumber \\ &\times&
\sixj{\kappa}{1/2}{1/2}{j_p}{J}{J'}
\sixj{\kappa}{J'}{J}{S}{\lambda}{\lambda'}
\sixj{l'}{l}{\kappa}{\lambda}{\lambda'}{L}
\nonumber \\ &\times &
\mathcal{M}(plPL\lambda SJT;{\jL \jp})
\mathcal{M}^*(pl'PL\lambda' SJ'T';{\jL \jp}) \,,
\label{omegas1}
\end{eqnarray}
with $L=0$ for the  $1s_{1/2}$ state, and  $L=0$ and $1$ for the  $1p_{3/2}$ state. 

It is interesting to observe that the presence of the Clebsch-Gordan coefficient in Eq.~\Ref{omegas1}, for $\kappa=1$, ensures that $l$ and $l'$ have opposite parities. Since the initial state in the two matrix elements has a definite parity, this implies that all contributions to $\omega_1$ come from interference terms between the parity-conserving and the parity-violating parts of the transition potential. Furthermore, the antisymmetrization factor in Eq.~\Ref{twobody} shows that the two final states have $T\neq~T'$. These are general properties of the asymmetry parameter, as mentioned in the introduction.

\section{Numerical results and conclusions \label{numerical} }

%
%
\begin{table}[htb]
\caption{Results for the asymmetry parameter, $a_\Lambda$, based on the nonmesonic decay of $^{5}_\Lambda$He. See text for detailed explanation.
\label{helium}
}
\smallskip
\begin{ruledtabular}
\begin{tabular}{|c|ccc|}
Model/Calculations &
$\omega_0(1s_{1/2})$ &
$\omega_1(1s_{1/2})$ &
$a_\Lambda$ \\
\hline
$\pi$&&&\\
Strictly local&$0.5176$&$-0.2254$&$-0.4354$ ($-0.4351$)\\ 
Plus corrections&$0.6492$&$-0.2913$&$-0.4487$ ($-0.4456$)\\  
&&&\\
$(\pi,\eta,K$)&&&\\
Strictly local&$0.3322$&$-0.1878$&$-0.5652$ ($-0.5852$)\\   
Plus corrections&$0.3920$&$-0.2412$&$-0.6153$ ($-0.6384$)\\  
&&&\\
$\pi+\rho$&&&\\
Strictly local&$0.5011$&$-0.1227$&$-0.2449$ ($-0.2665$)\\ 
Plus corrections&$0.5937$&$-0.1776$&$-0.2991$ ($-0.3155$)\\  
&&&\\
$(\pi,\eta,K)+(\rho,\omega,K^*)$&&&\\
Strictly local&$0.5352$&$-0.2739$&$-0.5117$ ($-0.5131$)\\  
Plus corrections&$0.5526$&$-0.2974$&$-0.5382$ ($-0.5388$)\\  
&&&\\
\hline
Experiment &
\multicolumn{2}{l}{KEK-PS~E278 \cite{Aj00}} &
$0.24 \pm 0.22$ \\
$a_\Lambda = A_V(^{5}_\Lambda\mathrm{He})$ &
\multicolumn{2}{l}{KEK-PS~E462 \cite{Ma05} (preliminary)} &
$0.11 \pm 0.08 \pm 0.04$ \\
\end{tabular}
\end{ruledtabular}
\end{table}
%

%
%
\begin{table}[htb]
\caption{Results for the asymmetry parameter, $a_\Lambda$, based on the nonmesonic decay of $^{12}_{\protect\phantom{1}\Lambda}$C. See text for detailed explanation.
\label{carbon}
}
\smallskip
\begin{ruledtabular}
\begin{tabular}{|c|ccccc|}
Model/Calculations &
$\omega_0(1s_{1/2})$ &
$\omega_0(1p_{3/2})$ &
$\omega_1(1s_{1/2})$ &
$\omega_1(1p_{3/2})$ &
$a_\Lambda$ \\
\hline
$\pi$&&&&&\\
Strictly local&$0.4111$&$0.4724$&$-0.1830$&$-0.1990$&$-0.4324$\\  
Plus corrections&$0.5206$&$0.5954$&$-0.2400$&$-0.2596$&$-0.4477$\\  
&&&&&\\
$(\pi,\eta,K$)&&&&&\\
Strictly local&$0.2788$&$0.3161$&$-0.1580$&$-0.1707$&$-0.5526$\\  
Plus corrections&$0.3336$&$0.3817$&$-0.2057$&$-0.2217$&$-0.5975$\\ 
&&&&&\\
$\pi+\rho$&&&&&\\
Strictly local&$0.4138$&$0.4607$&$-0.0984$&$-0.1096$&$-0.2379$\\  
Plus corrections&$0.4922$&$0.5514$&$-0.1461$&$-0.1596$&$-0.2930$\\  
&&&&&\\
$(\pi,\eta,K)+(\rho,\omega,K^*)$&&&&&\\
Strictly local&$0.4391$&$0.4803$&$-0.2300$&$-0.2378$&$-0.5088$\\  
Plus corrections&$0.4619$&$0.5083$&$-0.2546$&$-0.2599$&$-0.5303$\\  
&&&&&\\
\hline
Experiment &
\multicolumn{4}{l}{KEK-PS~E160 \cite{Aj92}} &
$-0.9 \pm 0.3$%
\footnote{This result corresponds to an improved weighted average among several $p$-shell hypernuclei \cite[p.95]{Al02}.}
\\
$a_\Lambda = -2 A_V(^{12}_{\protect\phantom{1}\Lambda}\mathrm{C})$ &
\multicolumn{4}{l}{KEK-PS~E508 \cite{Ma05} (preliminary)} &
$-0.44 \pm 0.32$%
\footnote{See text.} 
\\
\end{tabular}
\end{ruledtabular}
\end{table}

Shown in Tables \ref{helium} and \ref{carbon} are the results obtained in the calculation of the asymmetry parameter, $a_\Lambda$, based on the expressions of the previous section applied to $^5_\Lambda$He and $^{12}_{\phantom{1}\Lambda}$C, respectively. The values of $\omega_0$ and $\omega_1$ are in units of the free $\Lambda$ decay constant,   $\Gamma_\Lambda^{(0)}=2.50 \times 10^{-6}$ eV, and, in the case of $^{12}_{\phantom{1}\Lambda}$C, we give in separate columns the contributions of the $1s_{1/2}$ and $1p_{3/2}$ proton states. Also included are the measured values for $a_\Lambda$ obtained from some recent experiments on the nonmesonic decay of these two hypernuclei.
The value for $a_\Lambda$ in $^{12}_{\phantom{1}\Lambda}$C  extracted from experiment KEK-PS~E508 and given in Table~\ref{carbon} was taken from the preprint version of Ref.~\cite{Ma05}, since only a weighted average for $^{12}_{\phantom{1}\Lambda}$C and $^{11}_{\phantom{1}\Lambda}$B is explicitly reported in the published version, its value being $-0.20 \pm 0.26 \pm 0.04$. 

We consider several OME models, and for each one we give the results of two different calculations. First, are those corresponding to the strictly local approximation for the transition potential, usually adopted in the literature. Secondly, are the ones obtained when we add the corrections due to the kinematical effects related to the lambda-nucleon mass difference and the first-order nonlocality terms that we have discussed in Ref.~\cite{Ba03}.
The first thing to notice is that these corrections act systematically in the direction of increasing the absolute values of all the tabulated quantities. The effect is typically in the range of $20$--$30\%$ for $\omega_\kappa$ but only $5$--$10\%$ for $a_\Lambda$, tending to be larger in the $\pi+\rho$ model and smaller in the complete model. To put it shortly, if one wants precise values for the asymmetry parameter, the correction terms should be included in the transition potential, but, in view of the present level of indeterminacy in the measurements, they can be dispensed with for the moment.

In the case of $^5_\Lambda$He, we have also included, between parentheses, in
Table~\ref{helium}, the values for $a_\Lambda$ obtained with the approximate formula usually adopted in the literature \cite{Al02}, namely,
\begin{equation}\label{asym}
A_V(^5_\Lambda\mathrm{He}) \approx
\frac{2\,\Re \left[\sqrt{3}\, ae^* - b\, ( c^* - \sqrt{2}\, d^* ) +
\sqrt{3}\, f\, ( \sqrt{2}\, c^* + d^* ) \right]}
{ |a|^2 + |b|^2 + 3 \left( |c|^2 + |d|^2 + |e|^2 + |f|^2 \right) } \,,
\end{equation}
where
\begin{equation}\label{tramps}
\begin{array}[b]{lll}
a = \bra{np, ^1\!\mathrm{S}_0}V\ket{\Lambda p, ^1\!\mathrm{S}_0} \,,
\quad
&
b = i \bra{np, ^3\!\mathrm{P}_0}V\ket{\Lambda p, ^1\!\mathrm{S}_0} \,,
\quad
&
c = \bra{np, ^3\!\mathrm{S}_1}V\ket{\Lambda p, ^3\!\mathrm{S}_1} \,,
\\
& & \\
d = - \bra{np, ^3\!\mathrm{D}_1}V\ket{\Lambda p, ^3\!\mathrm{S}_1} \,,
\quad
&
e = i \bra{np, ^1\!\mathrm{P}_1}V\ket{\Lambda p, ^3\!\mathrm{S}_1} \,,
\quad
&
f = - i \bra{np, ^3\!\mathrm{P}_1}V\ket{\Lambda p, ^3\!\mathrm{S}_1} \,.
\end{array}
\end{equation}
The extra factors in the transition amplitudes in Eqs.~\Ref{tramps} are due to
differences in phase conventions, as explained in Ref.~\cite{Ba03}.
It is important to emphasize that Eq.~\Ref{asym} is only an approximation, that can  be adapted from the corresponding expression for the two-body reaction $pn \to p\Lambda$ in free space \cite{Na99}. As such, it ignores the fact that the final state of  nonmesonic decay is a three-body one and the ensuing kinematical complications should be properly dealt with, which requires a direct integration over the available phase space as done in the expressions used here. More importantly, Eq.~\Ref{asym} does not include the full contribution of the transitions coming from proton states beyond the $s$-shell, being therefore  of only limited validity, and should not be used for $p$-shell hypernuclei such as $^{12}_{\phantom{1}\Lambda}$C, or, even worse, for heavier ones. This being said, comparison of the corresponding values for $a_\Lambda$ in Table~\ref{helium} shows that the formula works well within its range of validity. This conclusion is in agreement with our preliminary result reported elsewhere \cite{Ba04}, which was restricted to one-pion-exchange only.

Coming now to $^{12}_{\phantom{1}\Lambda}$C, it is evident in Table~\ref{carbon} that the $p$-shell contributions to $\omega_0$ and $\omega_1$ are by no means negligible, being in fact of the same order as those of the $s$-shell. However, they are also in approximately the same ratio, so that the effect on $a_\Lambda$, given by Eq.~\Ref{parameter1}, is much smaller. This corroborates the theoretical expectation that the intrinsic asymmetry parameter, $a_\Lambda$, should have only a moderate dependence on the particular hypernucleus considered.
Presently we are investigating to which degree this remains true for more general cases, such as that of $^{11}_{\phantom{1}\Lambda}$B \cite{Ba04b}.
Notice that in the previous section we have explicitly proven that $a_\Lambda$ is independent of the hypernuclear spin.
However this does not, by itself, exclude the possibility that it might depend on other aspects of hypernuclear structure.

In closing, let us remark that we have derived simple formulas for the evaluation of the asymmetry parameter, which exactly include the effects of three-body kinematics in the final states of nonmesonic hypernuclear decay and correctly treat the contribution of transitions originating from proton states beyond the $s$-shell.
As to our numerical results, let us first of all observe that the calculated values of $a_\Lambda$ in the four OME models considered here vary from $-0.62$ to $-0.24$. This broad spectrum of values indicates that the asymmetry parameter can indeed be a powerful tool to discriminate between different theoretical mechanisms for nonmesonic decay, requiring for this purpose, however, a more precise experimental determination of this observable than those  presently available. Secondly, the fact that, for each of these OME models, the results for $^5_\Lambda$He and $^{12}_{\phantom{1}\Lambda}$C are very similar is compatible with the general expectation that $a_\Lambda$ should depend little on the hypernucleus. Finally, the negative value systematically obtained for $a_\Lambda$ for the two hypernuclei indicates, once again, that it will be hard to get a positive or zero value for it in the first case, at least within strict OME models. The puzzle posed by the experimental results for $a_\Lambda$ in $s$- and $p$-shell hypernuclei remains unexplained.
%
%
%
\begin{appendix}

\section{$K=2$ contributions to $\omega_0$ and $\omega_1$ \label{K2}}

As mentioned in Section~\ref{asympar}, to compute the transition matrix elements $\mathcal{M}$ appearing in Eq.~\Ref{omegas}, we perform a Moshinsky transformation of the initial $\Lambda\!\,p$ state \cite[Eq.(A.1)]{Ba03},
\begin{eqnarray}
\roundket{\jL \jp J}
&=& \hat{j}_\Lambda\, \hat{j}_p\;
\sum_{\bar{\lambda} \bar{S}}
\hat{\bar{\lambda}}\, \hat{\bar{S}}\;
\ninj{l_{\Lambda}}{1/2}{j_{\Lambda}}{l_p}{1/2}{j_p}
{\bar{\lambda}}{\bar{S}}{J}
\nonumber \\
&\times& \sum_{n\bar{l}NL}
\roundbraket{n\bar{l}\, NL\, \bar{\lambda}}
{n_{\Lambda} l_{\Lambda}\, n_p l_p\, \bar{\lambda}} \,
\roundket{n\bar{l}\,NL\, \bar{\lambda} \bar{S} J}
\nonumber \\
&\equiv& \sum_{n\bar{l}\,NL\, \bar{\lambda} \bar{S}}
C(n\bar{l}\,NL\, \bar{\lambda} \bar{S}; \jL \jp J) \,
\roundket{n\bar{l}\,NL\, \bar{\lambda} \bar{S} J} \,
\,,
\label{moshinsky}
\end{eqnarray}
where 
$\roundbraket{n\bar{l}\, NL\, \bar{\lambda}}
{n_{\Lambda} l_{\Lambda}\, n_p l_p\, \bar{\lambda}}$
are the Moshinsky brackets with their phases adapted so as to conform with our convention for the relative coordinate as discussed in Appendix A of Ref.~\cite{Ba03}. We have put bars over $\bar{l}$, $\bar{\lambda}$ and $\bar{S}$ to distinguish them from the analogous angular momenta in the partial waves of the final $N\!N$ state. This is not necessary for $L$ and $J$, since all transitions are diagonal in these two quantum numbers.
Introducing Eq.~\Ref{moshinsky} into Eq.~\Ref{twobody}, one gets
\begin{equation}\label{decomp}
\mathcal{M}(plPL\lambda SJT;\jL \jp) =
\sum_{n\bar{l}\,N\, \bar{\lambda} \bar{S}}
C(n\bar{l}\,NL\, \bar{\lambda} \bar{S}; \jL \jp J) \,
\mathcal{M}(plPL\lambda SJT; n\bar{l}\,N\, \bar{\lambda} \bar{S}) \,,
\end{equation}
where
\begin{eqnarray}
\M(plPL\lambda SJT;n\bar{l}\,N\, \bar{\lambda} \bar{S})
&=&\sqi\left[1-(-)^{l+S+T}\right]
\roundbra{plPL\lambda SJT}V\roundket{n\bar{l}\,NL\,\bar{\lambda}\bar{S}J} \,.
\label{twobody1}
\end{eqnarray}

The transition potential can be decomposed as
\begin{equation}\label{Vdecomp}
V = \sum_i v_i(r)\,
I_i\, \Omega_i \,,
\end{equation}
where the isospin factor $I_i$ is equal to $1$ or $\bm{\tau}_1 \cdot \bm{\tau}_2$, for isoscalar or isovector interactions, respectively, and the $\Omega_i$ are rotationally invariant operators having definite spin and spatial ranks, \ie, operators of the form
\begin{equation}\label{tensorform}
\Omega_i = \left[A_i^{\nu_i}(\bm{\sigma}_1, \bm{\sigma}_2) \otimes B_i^{\nu_i}(\bm{r},\nabla)\right]_{00} \,.
\end{equation}
Due to the algebraic properties of the Pauli matrices, the rank $\nu_i$ can be at most equal to 2. For the OME models we consider, including eventual kinematical and nonlocality corrections, the several possibilities are:
\begin{equation}\label{ranks}
\begin{array}[b]{rl}
\mbox{PC terms} & \left\{
\begin{array}{rl}
\nu_i = 0 &\quad \mbox{for central and  $\bm{r}\cdot\nabla$ forces}
\\
 &\quad \mbox{(spin-independent or spin-spin)}\,,
\\
\nu_i = 1 &\quad \mbox{for linear spin-orbit forces}\,,
\\
\nu_i = 2 &\quad \mbox{for tensor forces}\,;
\end{array}
\right.
\\ &
\\
\mbox{PV terms}  & \,\,\left\{
\begin{array}{rl}
\nu_i = 1 &\quad \mbox{for all kinds}\,.
\end{array}
\right.
\end{array}
\end{equation}
As can be seen in Eqs. (A.3) to (A.15) of Ref.~\cite{Ba03}, the different terms have matrix elements of the general form
\begin{equation}
\roundbra{plPL\lambda SJT}v_iI_i\Omega_i\roundket{n\bar{l}\,NL\,\bar{\lambda}\bar{S}J} =
G_i(lL\lambda SJT; \bar{l} \bar{\lambda} \bar{S})\,
({PL}|{NL})\,
({pl}|v_i(r)\hat{d}_i(r)|{n\bar{l}}) \,,
\end{equation}
where $({PL}|{NL})$ are the overlaps of the c.m.~radial wave functions and the $\hat{d}_i(r)$ are, either unity, or one of the effective differential operators defined in Eq.~(A.16) of that reference. The important point is that the $G_i$ are purely geometrical factors, involving, at most, $3j$ and $6j$ symbols.
Scrutinizing these equations more closely, one notices that the dependence on $\lambda$, $\bar{\lambda}$ and $J$ can be isolated as follows
\begin{equation}\label{geometric}
G_i(lL\lambda SJT; \bar{l} \bar{\lambda} \bar{S}) =
(-)^J\,
\hat{\lambda}\hat{\bar{\lambda}}\,
\sixj{\lambda}{\nu_i}{\bar{\lambda}}{\bar{l}}{L}{l}\,
\sixj{\lambda}{\nu_i}{\bar{\lambda}}{\bar{S}}{J}{S}\,
g_i(lLST; \bar{l} \bar{S}) \,.
\end{equation}
This result is completely general and depends only on the application of the Wigner-Eckart theorem to operators of the form given in Eq.~\Ref{tensorform}. (See, for instance, Chapter 7 of Ref.~\cite{Ed74}, or Section 1A-5 of Ref.~\cite{Bo69}.)

Taking these ideas into account in Eq.~\Ref{omegas}, it is clear that the summation over $\lambda$, $J$, $\lambda'$ and $J'$ can be performed first, leading to a remaining summand that is proportional to
\begin{eqnarray}
\lefteqn{
\sum_{\lambda J} \; \sum_{\lambda' J'}
(-)^{\lambda+J+J'}\;
\hat{\lambda}^2 \hat{\lambda'}^2 \hat{J}^2 \hat{J'}^2
}
\nonumber \\ &\times&
\sixj{\kappa}{1/2}{1/2}{j_p}{J}{J'}
\sixj{\kappa}{J'}{J}{S}{\lambda}{\lambda'}
\ninj{l}{l'}{k}{L}{L'}{K}{\lambda}{\lambda'}{\kappa}
\nonumber \\ &\times&
\ninj{l_{\Lambda}}{1/2}{j_{\Lambda}}{l_p}{1/2}{j_p}
{\bar{\lambda}}{\bar{S}}{J}
\sixj{\lambda}{\nu_i}{\bar{\lambda}}{\bar{l}}{L}{l}\,
\sixj{\lambda}{\nu_i}{\bar{\lambda}}{\bar{S}}{J}{S}\,
\nonumber \\ &\times&
\ninj{l_{\Lambda}}{1/2}{j_{\Lambda}}{l_p}{1/2}{j_p}
{\bar{\lambda}'}{\bar{S}'}{J'}
\sixj{\lambda'}{\nu_{i'}}{\bar{\lambda}'}{\bar{l}'}{L'}{l'}\,
\sixj{\lambda'}{\nu_{i'}}{\bar{\lambda}'}{\bar{S}'}{J'}{S}\,.
\label{coeff0}
\end{eqnarray}
Actually, the $\Lambda$ is always in the $1s_{1/2}$ state, and one can make use of Eq.~(6.4.14) of Ref.~\cite{Ed74} to replace the above expression by
\begin{eqnarray}
X &\equiv&
\sum_{\lambda J} \; \sum_{\lambda' J'}
(-)^{\lambda+J+J'}\;
\hat{\lambda}^2 \hat{\lambda'}^2 \hat{J}^2 \hat{J'}^2
\nonumber \\ &\times&
\sixj{\kappa}{1/2}{1/2}{j_p}{J}{J'}
\sixj{\kappa}{J'}{J}{S}{\lambda}{\lambda'}
\ninj{l}{l'}{k}{L}{L'}{K}{\lambda}{\lambda'}{\kappa}
\nonumber \\ &\times&
\sixj{\bar{S}}{1/2}{1/2}{j_p}{l_p}{J}\,
\sixj{\lambda}{\nu_i}{l_p}{\bar{l}}{L}{l}\,
\sixj{\lambda}{\nu_i}{l_p}{\bar{S}}{J}{S}\,
\nonumber \\ &\times&
\sixj{\bar{S}'}{1/2}{1/2}{j_p}{l_p}{J'}\,
\sixj{\lambda'}{\nu_{i'}}{l_p}{\bar{l}'}{L'}{l'}\,
\sixj{\lambda'}{\nu_{i'}}{l_p}{\bar{S}'}{J'}{S}\,,
\label{coeff}
\end{eqnarray}
where we have dropped an irrelevant factor.

For a proton in the $s$ shell, the Moshinsky transformation requires that $L=L'=0$, and the $9j$ symbol in Eq.~\Ref{coeff} selects $K=0$ as the only possibility. 
For a proton in the $p$ shell, there are two alternatives for the relative and c.m.~angular momenta, namely,
\begin{equation}\label{pshell}
\mbox{$p$ shell } \left\{ 
\begin{array}{rl}
\mbox{alternative 1: }  &  \bar{l} = 1 \mbox{ and } L=0 \,,
\\
\mbox{alternative 2: }  & \bar{l} = 0 \mbox{ and } L=1 \,,
\end{array}
\right.
\end{equation}
and similarly for the primed quantities.
In principle, therefore, there are three possibilities for $K$, namely, $K=0,1$ and $2$.

It is clear from this discussion that a contribution with $K=2$ in Eq.~\Ref{omegas} can only come from the alternative 2 in Eq.~\Ref{pshell}. Therefore, setting $L=L'=1$ and $\bar{l}=\bar{l}'=0$ in  Eq.~\Ref{coeff}, and making use of Eq.~(6.3.2) of Ref.~\cite{Ed74}, we get 
\begin{eqnarray}
\lefteqn{
X(\mathsf{j}_p=1p_{j_p};\; L=L'=1)
}
\nonumber \\ &=&
\frac{(-)^{l+l'}}{3\hat{l}\hat{l}'}\,\delta_{\nu_{i} l}\, \delta_{\nu_{i'} l'}\,
\sum_{\lambda J} \; \sum_{\lambda' J'}
(-)^{\lambda'+J+J'}\;
\hat{\lambda}^2 \hat{\lambda'}^2 \hat{J}^2 \hat{J'}^2
\nonumber \\ &\times&
\sixj{\kappa}{1/2}{1/2}{j_p}{J}{J'}
\sixj{\kappa}{J'}{J}{S}{\lambda}{\lambda'}
\ninj{l}{l'}{k}{1}{1}{K}{\lambda}{\lambda'}{\kappa}
\nonumber \\ &\times&
\sixj{\bar{S}}{1/2}{1/2}{j_p}{1}{J}\,
\sixj{\lambda}{l}{1}{\bar{S}}{J}{S}\,
\nonumber \\ &\times&
\sixj{\bar{S}'}{1/2}{1/2}{j_p}{1}{J'}\,
\sixj{\lambda'}{l'}{1}{\bar{S}'}{J'}{S}\,.
\label{coeff1}
\end{eqnarray}

To proceed, it will be unavoidable to perform some manipulations with $12j$ symbols, and  the needed identities are collected in Appendix~\ref{12j} for convenience. 
With the help of well known symmetry properties of $6j$ and $9j$ symbols \cite{Ed74}, one can make use of Eqs. \Ref{recursion}, \Ref{symmetry} and \Ref{reduction}, in succession, to perform, first the summation over $\lambda$, and then that over $\lambda'$, in Eq.~\Ref{coeff1}, getting
\begin{eqnarray}
\lefteqn{
X(\mathsf{j}_p=1p_{j_p};\; L=L'=1)
}
\nonumber \\ &=&
\frac{(-)^{K+l'+\bar{S}'}}{3\hat{l}\hat{l}'}\,\delta_{\nu_{i} l}\, \delta_{\nu_{i'} l'}\,
\sixj{k}{\bar{S}}{\bar{S}'}{S}{l'}{l}
\nonumber \\ &\times&
\sum_{J} \; \sum_{J'}
(-)^{J'}\;
\hat{J}^2 \hat{J'}^2 \,
\ninj{k}{\bar{S}}{\bar{S}'}{\kappa}{J}{J'}{K}{1}{1}
\nonumber \\ &\times&
\sixj{\kappa}{1/2}{1/2}{j_p}{J}{J'}\,
\sixj{\bar{S}}{1/2}{1/2}{j_p}{1}{J}\,
\nonumber \\ &\times&
\sixj{\bar{S}'}{1/2}{1/2}{j_p}{1}{J'} \,.
\label{coeff2}
\end{eqnarray}
Repeating the same procedure, we can now perform, first the summation over $J$, and then that over $J'$, to get
\begin{eqnarray}
\lefteqn{
X(\mathsf{j}_p=1p_{j_p};\; L=L'=1)
}
\nonumber \\ &=&
- \frac{(-)^{K+k+l'+\bar{S}'}}{3\hat{l}\hat{l}'}\,
\delta_{\nu_{i} l}\, \delta_{\nu_{i'} l'}\,
\sixj{k}{\bar{S}}{\bar{S}'}{S}{l'}{l}
\nonumber \\ &\times&
\ninj{K}{1/2}{1/2}{\kappa}{1/2}{1/2}{k}{\bar{S}}{\bar{S}'}\,
\sixj{K}{1/2}{1/2}{j_p}{1}{1} \,.
\label{coeff3}
\end{eqnarray}

The $9j$, as well as the last $6j$, in Eq.~\Ref{coeff3} restricts $K$ to $0$ and $1$, and we conclude that the contribution with $K=2$ in Eq.~\Ref{omegas} vanishes identically, both for $\kappa=0$, and for $\kappa=1$.
Notice that this result holds, not only for $\jp=1p_{3/2}$, which is of direct interest for $^{12}_{\phantom{1}\Lambda}$C, but also for $\jp=1p_{1/2}$, which may be relevant for other $p$-shell hypernuclei.

\section{$K=1$ contribution to $\omega_0$ \label{K10}}

Recalling Eq.~\Ref{compact2}, it is clear that the 
$K=1$ contribution to $\omega_0$ in Eq.~\Ref{omegas},  from the single-proton state $\jp$, 
has the form 
\begin{eqnarray}
\lefteqn{
\omega_0(\jp, \, K=1) 
} 
\nonumber \\ &=& 
\int d\cos\theta_{p_1}\, 
\int p_2^2\, dp_2 \int p_1^2\, dp_1\;
\delta\left( \frac{p_1^2}{2\Mass} + \frac{p_2^2}{2\Mass}
+ \frac{|\bm{p}_1 + \bm{p}_2|^2}{2\Mass_F}
- \Delta_{\jp} \right)
\nonumber \\ &\times& f(p,P) 
[Y_{1}(\theta_p,\pi)\otimes Y_{1}(\theta_P,0)] _{0 0} \,,
\label{K10a}
\end{eqnarray} 
where $f(p,P)$ represents the rest of the integrand in Eq.~\Ref{omegas}, the important point being that it depends on the momenta only through $p$ and $P$. 

From the explicit expressions of the spherical harmonics, we find 
\begin{equation}
[Y_{1}(\theta_p,\pi)\otimes Y_{1}(\theta_P,0)] _{0 0} = 
- \frac{\sqrt{3}}{4\pi} \cos (\theta_p + \theta_P) \,,
\end{equation}
and, making use of the last two equations in \Ref{variables}, this becomes 
\begin{equation}
[Y_{1}(\theta_p,\pi)\otimes Y_{1}(\theta_P,0)] _{0 0} = 
- \frac{\sqrt{3}}{4\pi}\, \frac{p_2^2 - p_1^2}{2pP} \,.
\end{equation} 
Introducing this result in Eq.~\Ref{K10a}, we are left with 
\begin{eqnarray}
\lefteqn{
\omega_0(\jp, \, K=1) 
} 
\nonumber \\ &=& 
- \frac{\sqrt{3}}{4\pi}\,
\int d\cos\theta_{p_1}\, 
\int p_2^2\, dp_2 \int p_1^2\, dp_1\;
\delta\left( \frac{p_1^2}{2\Mass} + \frac{p_2^2}{2\Mass}
+ \frac{|\bm{p}_1 + \bm{p}_2|^2}{2\Mass_F}
- \Delta_{\jp} \right)
\nonumber \\ &\times& f(p,P) \, \frac{p_2^2 - p_1^2}{2pP} \,.
\label{K10b}
\end{eqnarray} 
Let us now perform the interchange of dummy variables $p_1 \leftrightarrow p_2$ in Eq.~\Ref{K10b}, keeping $\cos\theta_{p_1}$ fixed.  Then, noticing that, according to the first two equations in \Ref{variables}, $p$ and $P$ are invariant under this transformation, we arrive at the result 
\begin{equation}
\omega_0(\jp, \, K=1) = {} - \omega_0(\jp, \, K=1) \,,
\end{equation}
from which it follows that the contribution with $K=1$ in Eq.~\Ref{omegas} vanishes for $\kappa=0$. 

\section{$K=1$ contribution to $\omega_1$ \label{K11}}

%
%
\begin{table}[htb]
\caption{Results for the $K=1$ contribution to $\omega_1$ in the nonmesonic decay of $^{12}_{\protect\phantom{1}\Lambda}$C. See text for detailed explanation.
\label{Keqone}
}
\smallskip
\begin{ruledtabular}
\begin{tabular}{|c|cc|}
Model/Calculations &
$\omega_1(1p_{3/2},\, K=1)$ &
$|\Delta a_\Lambda/a_\Lambda|$ (\%) \\
\hline
$\pi$&&\\
Strictly local&$0.0007$&$0.18$\\ 
Plus corrections&$0.0010$&$0.20$\\  
&&\\
$(\pi,\eta,K$)&&\\
Strictly local&$0.0005$&$0.15$\\   
Plus corrections&$0.0007$&$0.16$\\  
&&\\
$\pi+\rho$&&\\
Strictly local&$0.0007$&$0.34$\\  
Plus corrections&$0.0008$&$0.26$\\  
&&\\
$(\pi,\eta,K)+(\rho,\omega,K^*)$&&\\
Strictly local&$0.0003$&$0.06$\\  
Plus corrections&$0.0006$&$0.12$\\  
&&\\
\end{tabular}
\end{ruledtabular}
\end{table}

We have not been able to find an analytical proof that the $K=1$ contribution in Eq.~\Ref{omegas} vanishes also for $\kappa=1$. However, for the cases we are dealing with, this contribution can only arise from the $\jp=1p_{3/2}$ proton state in $^{12}_{\phantom{1}\Lambda}\mathrm{C}$, and we have numerically computed its values in the several  OME models we are considering. They are given, in units of $\Gamma_\Lambda^{(0)}$, in Table~\ref{Keqone}.  
The nonzero values may be due to truncation and roundoff errors. For instance, we have computed,  with the same routine, the analogous contribution with $\kappa=0$ in the case of the complete OME model plus kinematical and nonlocality terms. Even though it has been proved in Appendix~\ref{K10} that this is exactly zero, the numerical result came out as 0.0004, which is comparable to the value obtained for $\omega_1(1p_{3/2},\, K=1)$ in the same model. 

Comparison of Tables \ref{carbon} and \ref{Keqone} immediately shows that 
the $K=1$ contribution to $\omega_1$ is, in any case, very small. Furthermore, in the last column of Table~\ref{Keqone}, we give the relative effect that its   inclusion would have on $a_\Lambda$, and it always stays below $0.4\%$. We conclude that, even if this contribution is not exactly zero, it can be safely neglected. 

\section{Some properties of $12j$ symbols \label{12j}}

The $12j$ symbols arise in the recoupling of five angular momenta \cite{Ro59,Yu62}. They are not unique, but here we shall need only those of the first kind,
\begin{eqnarray}
\lefteqn{
\twlvj{\row{j_1}{j_2}{j_3}{j_4}}{\row{l_1}{l_2}{l_3}{l_4}}{\row{k_1}{k_2}{k_3}{k_4}}
=
\sum_{x} (-)^{R_4 - x} \, \hat{x}^2 \,
}
\nonumber \\
&\times&
\sixj{j_1}{k_1}{x}{k_2}{j_2}{l_1} \,
\sixj{j_2}{k_2}{x}{k_3}{j_3}{l_2} \,
\sixj{j_3}{k_3}{x}{k_4}{j_4}{l_3} \,
\sixj{j_4}{k_4}{x}{j_1}{k_1}{l_4} \,,
\label{definition}
\end{eqnarray}
as defined in Eq.~(19.1) of Ref.~\cite{Yu62}, whose notation for the $12j$ symbols we follow.
In Eq.~\Ref{definition}, $R_4$ stands for the sum of all the angular momentum labels in the $12j$ symbol.

These symbols obey the recursion relation \cite[Eq.(A.6.13)]{Yu62}
\begin{eqnarray}
\lefteqn{
\twlvj{\row{j_1}{j_2}{j_3}{j_4}}{\row{l_1}{l_2}{l_3}{l_4}}{\row{k_1}{k_2}{k_3}{k_4}}
=
(-)^{j_2 + k_2 + j_4 + k_4} \,
}
\nonumber \\
&\times&
\sum_{x} (-)^{2x} \, \hat{x}^2 \,
\ninj{k_2}{j_4}{x}{l_2}{j_3}{j_2}{k_3}{l_3}{k_4} \,
\sixj{k_2}{j_4}{x}{l_4}{l_1}{k_1} \,
\sixj{k_4}{j_2}{x}{l_1}{l_4}{j_1} \,,
\label{recursion}
\end{eqnarray}
and have several symmetry properties, among which \cite[Eq.(17.4)]{Yu62}
\begin{equation} \label{symmetry}
\twlvj{\row{j_1}{j_2}{j_3}{j_4}}{\row{l_1}{l_2}{l_3}{l_4}}{\row{k_1}{k_2}{k_3}{k_4}}
=
\twlvj{\row{j_2}{j_3}{j_4}{k_1}}{\row{l_2}{l_3}{l_4}{l_1}}{\row{k_2}{k_3}{k_4}{j_1}} \,.
\end{equation}
There are also reduction formulas, such as \cite[Eq.(A.6.39)]{Yu62}
\begin{eqnarray}
\lefteqn{
\sum_{x} \, \hat{x}^2 \,
\twlvj{\row{j_1}{j_2}{j_3}{j_4}}{\row{l_1}{l_2}{l_3}{l_4}}{\row{x}{k_2}{k_3}{k_4}} \,
\sixj{l_1}{k_2}{x}{j_4}{l_4}{k_1}
}
\nonumber \\
&=&
(-)^{l_2 + l_3 + j_3 + k_3 - k_1} \,
\ninj{j_2}{k_4}{k_1}{j_3}{l_3}{j_4}{l_2}{k_3}{k_2} \,
\sixj{j_2}{k_4}{k_1}{l_4}{l_1}{j_1} \,.
\label{reduction}
\end{eqnarray}
\end{appendix}
%
%

%

%

\begin{thebibliography}{99}
%
\bibitem{Al02}
W. M. Alberico and G. Garbarino, Phys. Rep. \textbf{369}, 1 (2002).
\bibitem{Aj92}
S. Ajimura \emph{et al.}, Phys. Lett. B \textbf{282}, 293 (1992).
\bibitem{Aj00}
S. Ajimura \emph{et al.}, Phys. Rev. Lett. \textbf{84}, 4052 (2000).
\bibitem{Ma05}
T. Maruta \emph{et al.}, Nucl. Phys. \textbf{A754}, 168c (2005). 
See also  \texttt{nucl-ex/0402017}.
\bibitem{Du96}
J. F. Dubach, G. B. Feldman, B. R. Holstein and L. de la Torre, Ann. Phys.
(N.Y.) \textbf{249}, 146 (1996).
\bibitem{Pa97}
A. Parre\~no, A. Ramos and C. Bennhold, Phys. Rev. C \textbf{56}, 339 (1997);
A. Parre\~no and A. Ramos, Phys. Rev. C \textbf{65}, 015204 (2002).
\bibitem{Ba02}
C. Barbero, D. Horvat, F. Krmpoti\'{c}, T. T. S. Kuo, Z. Naran\v{c}i\'c
and D. Tadi\'{c}, Phys. Rev. C \textbf{66}, 055209 (2002).
\bibitem{Ba03}
C. Barbero, C. De Conti, A. P. Gale\~ao and F. Krmpoti\'c, Nucl. Phys.
\textbf{A726}, 267 (2003).
\bibitem{It02}
K. Itonaga, T. Ueda and T. Motoba, Phys. Rev. C \textbf{65}, 034617 (2002);
K. Itonaga, T. Motoba and T. Ueda, \emph{Electrophoto Production of Strangeness on Nucleons and Nuclei} (Sendai03), K. Maeda, H. Tamura, S.N. Nakamura and O. Hashimoto eds., World Scientific (2004) pp. 397--402.
\bibitem{Sa02}
K. Sasaki, T. Inoue and M. Oka, Nucl. Phys.  \textbf{A707}, 477 (2002);
\emph{ibid} \textbf{A669}, 331 (2000); Erratum, \emph{ibid} \textbf{A678}, 455 (2000).
\bibitem{Ga04}
G. Garbarino, A. Parre\~no and A. Ramos, Phys. Rev. C \textbf{69}, 054603 (2004) and references therein.
\bibitem{Al05}
W. M. Alberico, G. Garbarino, A. Parre\~no and A. Ramos, Phys. Rev. Lett. \textbf{94}, 082501 (2005). 
\bibitem{Ra92}
A. Ramos, E. van Meijgaard, C. Bennhold and B.K. Jennings, Nucl. Phys.
\textbf{A544}, 703 (1992).
\bibitem{Pa04}
A. Parre\~no, C. Bennhold and B. R. Holstein, Phys. Rev. C \textbf{70}, 051601(R) (2004).
\bibitem{Sa05}
K. Sasaki, M. Izaki and M. Oka, Phys. Rev. C \textbf{71}, 035502 (2005). 
\bibitem{Ej87}
H. Ejiri, T. Fukuda, T. Shibata, H. Band\={o} and K.-I. Kubo, Phys. Rev. C \textbf{36}, 1435 (1987).
\bibitem{Au70}
N. Austern, \emph{Direct Nuclear Reaction Theories}, Wiley-Interscience, N.Y., 1970.
\bibitem{Ga04a}
A. P. Gale\~ao, in preparation.
\bibitem{Ga04b}
A. P. Gale\~ao, in \textit{IX Hadron Physics and VII Relativistic Aspects of Nuclear Physics: A Joint Meeting on QCD and QGP}, Rio de Janeiro, 28 March -- 3 April 2004, edited by M. E. Bracco, M. Chiapparini, E. Ferreira, and T. Kodama (AIP Conference Proceedings \textbf{739}, 2004), pp. 560 -- 562. See also extended version mentioned therein.
\bibitem {Kr03}
F. Krmpoti\'c and  D. Tadi\'c, Braz. J. Phys. {\bf 33}, 187 (2003).
\bibitem{Mo59}
M. Moshinsky, Nucl. Phys. \textbf{13}, 104 (1959).
\bibitem{Na99}
H. Nabetani, T. Ogaito, T. Sato and T. Kishimoto, Phys. Rev. C \textbf{60}, 017001
(1999).
\bibitem{Ba04}
C. Barbero, A. P. Gale\~ao and F. Krmpoti\'c, Braz. J. Phys. \textbf{34}, 822 (2004).
\bibitem{Ba04b}
C. Barbero, A. P. Gale\~ao and F. Krmpoti\'c, work in progress.
\bibitem{Ed74}
A. R. Edmonds, \emph{Angular Momentum in Quantum Mechanics}, Princeton University Press, Princeton, N.J., 1974.
\bibitem{Bo69}
A. Bohr and B. R. Mottelson, \emph{Nuclear Structure}, Vol. I, W. A. Benjamin, Inc., New York, 1969.
\bibitem{Ro59}
M. Rotenberg, R. Bivins, N. Metropolis and J. K. Wooten, Jr., \emph{The 3-j and 6-j Symbols}, The Technology Press, MIT, Cambridge, MA, 1959.
\bibitem{Yu62}
A. P. Yutsis, I. B. Levinson and V. V. Vanagas, \emph{Mathematical Apparatus of the Theory of Angular Momentum}, published for the National Science Foundation, Washington,  D.C., by the Israel Program for Scientific Translations, Jerusalem, 1962. Translated from Russian by A. Sen and R. N. Sen.
%
\end{thebibliography}
\end{document}